# Nonlinear interactions between the Amazon River basin and the Tropical North Atlantic at interannual timescales

Alejandro Builes-Jaramillo[1, 2], Norbert Marwan[3], Germán Poveda[1] and Jürgen Kurths[3].

[1] Universidad Nacional de Colombia, Sede Medellín, Department of Geosciences and Environment, Facultad de Minas, Carrera 80 No 65-223, Bloque M2. Medellín, Colombia.

[2] Institución Universitaria Colegio Mayor de Antioquia, Facultad de Arquitectura e Ingeniería, Carrera 78 # 65 – 46, Edificio patrimonial. Medellín, Colombia.

[3] Potsdam Institute for Climate Impact Research, P. O. Box 6012 03, 14412 Potsdam, Germany

**Corresponding author:**

Alejandro Builes-Jaramillo
luis.builes@colmayor.edu.co
Telephone (+57) 3012422399

**Abstract**

We study the physical processes involved in the potential influence of Amazon (AM) hydroclimatology over the Tropical North Atlantic (TNA) Sea Surface Temperatures (SST) at interannual timescales, by analyzing time series of the precipitation index (P-E) over AM, as well as the surface atmospheric pressure gradient between both regions, and TNA SSTs. We use a recurrence joint probability based analysis that accounts for the lagged nonlinear dependency between time series, which also allows quantifying the statistical significance, based on a twin surrogates technique of the recurrence analysis. By means of such nonlinear dependence analysis we find that at interannual timescales AM hydrology influences future states of the TNA SSTs from 0 to 2 months later with a 90% to 95% statistical confidence. It also unveils the existence of two-way feedback mechanisms between the variables involved in the processes: (i) precipitation over AM leads the atmospheric pressure gradient between TNA and AM from 0 and 2 month lags, (ii) the pressure gradient leads the trade zonal winds over the TNA from 0 to 3 months and from 7 to 12 months, (iii) the zonal winds lead the SSTs from 0 to 3 months, and (iv) the SSTs lead precipitation over AM by 1 month lag. The analyses were made for time series spanning from 1979 to 2008, and for extreme precipitation events in the AM during the years 1999, 2005, 2009 and 2010. We also evaluated the monthly mean conditions of the relevant variables during the extreme AM droughts of 1963, 1980, 1983, 1997, 1998, 2005, and 2010, and also during the floods of 1989, 1999, and 2009. Our results confirm that the Amazon River basin acts as a land surface-atmosphere bridge that links the Tropical Pacific and TNA SSTs at interannual timescales. The identified mutual interactions between TNA and AM are of paramount importance for a deeper understanding of AM hydroclimatology but also of a suite of oceanic and atmospheric phenomena over the TNA, including recently observed trends in SSTs, as well as future occurrences and impacts on tropical storms and hurricanes throughout the TNA region, but also on fires, droughts, deforestation and dieback of the tropical rain forest of the Amazon River basin.

## 1. Introduction

Understanding the space-time dynamics of sea surface temperatures (SST) over the Tropical North Atlantic (TNA) is utterly important , given that this particular region has the potential to strongly modulate tropical and extra tropical climates (Avissar and Werth 2004). There is evidence that the TNA SSTs play a key role in the modulation of extreme hydrometeorological events in the boreal summer, such as droughts in the United States and heat waves in Europe (Sutton and Hodson 2005). Tropical Atlantic SSTs above 26°C in the TNA are one of the necessary conditions for the development of tropical storms and hurricanes reaching the Caribbean Sea, Central America, the US and the mid-latitude North Atlantic (Goldenberg et al. 2001; Trenberth 2005; Trenberth and Shea 2006; Chen et al. 2015). Even more, warm SSTs anomalies in the TNA region have been suggested as a trigger of El Niño events (Ham et al. 2013; Wang et al. 2017).

Changes in TNA SSTs are also determinant for the availability of moisture in the Amazon River basin, since the dynamics of the Intertropical Convergence Zone (ITZC) follows the annual cycle of SST over the TNA, which is associated with well-known precipitation patterns over land (Aceituno 1988; Marengo 1992; Fu et al. 2001; Poveda et al. 2006; Nobre et al. 2009; Yoon and Zeng 2010; Gimeno et al. 2012; Yin et al. 2014; Arias et al. 2015). The activity of the trade winds over the TNA region is related with the South American Low-level Jet (SALLJ), the South American Monsoon System (SAMS), and the dynamics of aerial rivers that combined configure the transport of moisture from Amazonia all the way to the eastern flank of the Andes, and upon veering to the southeast reach as far as northern Argentina (Marengo et al. 2004; Vera et al. 2006; Moraes-Arraut et al. 2011; Poveda et al. 2014). Also, TNA SSTs can modulate the interannual variability of fires in southern Amazonia (Chen et al. 2011; Fernandes et al. 2011).

The dynamics of SSTs in the TNA are controlled by diverse mechanisms acting at multidecadal, interannual, seasonal and intra-seasonal timescales, such as the Atlantic Multidecadal Oscillation (AMO) that operates on time scales of more than 50 years (Steinman et al. 2015), and the North-South Atlantic SST dipole that operates on time scales of 8-12 years (Enfield and Mayer 1997; Enfield et al. 1999). At interannual timescales there is a strong influence of El Niño on TNA SSTs positive anomalies, resulting, among others, from an anomalous Walker circulation that drives the teleconnection between the tropical Pacific and Atlantic basins (Saravanan and Chang 2000; Chiang and Sobel 2002), while the study by Poveda and Mesa (1997) showed statistical evidence and concluded that such interaction is mediated by land surface-atmosphere feedbacks and convection anomalies over the Amazon River basin, leading to the warming of SSTs over the TNA region following an El Niño event. Those authors went on to propose that tropical South America acts as a land surface-atmosphere bridge connecting the tropical Pacific and TNA.

On the other hand, the hydrometeorology of the Amazon River basin is not a passive spectator receiving the influence from large-scale ocean-atmosphere phenomena. There is evidence of a systematic change in the ITZC location during the boreal spring related with the propagation of coupled convective Kelvin waves emerging from AM and travelling towards the oceanic region (Wang and Fu 2007). Also, wet anomalies in the equatorial Amazon have been related with SST negative anomalies over the Caribbean and the Gulf of Mexico (Misra and DiNapoli 2012) at seasonal time scales.

Through lagged correlations, the work of Yoon and Zeng (2010) evidenced a possible feedback mechanism between AM precipitation and TNA SSTs, such that the decrease of convective activity in the Amazon leads the increase of TNA SSTs with a 3 to 5 month-lag, although those authors discarded any feedbacks and interpreted it in terms of the influence of the Pacific Ocean over both the AM and TNA regions. We are interested in further investigating the possible influence of the AM over the TNA SSTs and detailing the existence of two-way feedback mechanisms acting between the AM and TNA regions at interannual timescales. Our analysis is based on previous studies such Poveda and Mesa (1997), as well as Misra and DiNapoli (2012), on the influence of the AM precipitation over the Caribbean SSTs.

We aim at investigating the mechanisms proposed by Poveda and Mesa (1997) to provide further support to the existence of feedback processes between the hydroclimatology of the Amazon River basin and ocean-atmosphere processes over the TNA. To that end, we use a dynamical systems approach that overcomes diverse well-known shortcomings of cross-correlation analysis, in particular the highly non-linear character of the processes involved, which may introduce spurious cross-correlations with other hydro-climatic variables, thus clouding the interpretation of the physical mechanisms dynamics (Olden and Neff 2001; Runge et al. 2014).

We use a recurrence measure to quantify the non-linear lagged dependence based on the evolution of trajectories of the dynamical system in phase space, which does not require an *a priori* knowledge of the equations governing the system's behavior, useful in the analysis of hydrological and climatological time series (Marwan and Kurths 2002; Panagoulia and Vlahogianni 2014), and to unveil the directionality and time delay in the relation between two or more variables (Goswami et al. 2013; Marwan et al. 2013). Recurrence analysis does not require fitting of any kind of model to data, thus providing an alternative for studying non-linear, non-stationary, and high-dimensional processes (Proulx et al. 2009).

The present work is organized as follows. Section 2 outlines the main physical hypothesis of the feedback mechanism between the two regions and the data used in the present analysis. Section 3 provides a detailed methodological description of the recurrence framework. In section 4 we present the results of the recurrence analysis, and finally section 5 presents the concluding remarks and implications of the results.

## 2. The Proposed Mechanism and Data Sets

Here we propose the following physical processes and variables involved in the proposed two-way feedback mechanisms between the AM and TNA regions at interannual timescales (Fig. 1a). Driest (wettest) events in Amazonia (represented by the Precipitation Index, *P-E*), hereafter denoted as *P*, are characterized by a decrease (increase) in rainfall, which in turn acts to increase (decrease) atmospheric surface pressure in the AM region, thus reducing (increasing) the surface atmospheric pressure gradient between the TNA and AM regions, denoted as *G*, which in turn causes a slowing down (speeding up) of the zonal trade wind velocities over the TNA region, denoted as *W* (Fig. 1b). The weakening (strengthening) of the trade winds over the TNA is associated with a reduction (increase) in evaporative cooling and a subsequent increase (reduction) in TNA SSTs, denoted as *S*. Fig 2 shows results from lagged cross-correlation analysis between the time series of the *P-E* Index and TNA SSTs at interannual timescales (filtering out the annual cycles), which indicates that the AM hydrology leads the future TNA SSTs (Fig. 1b) from 0 to 2 month-lags (positive lags).

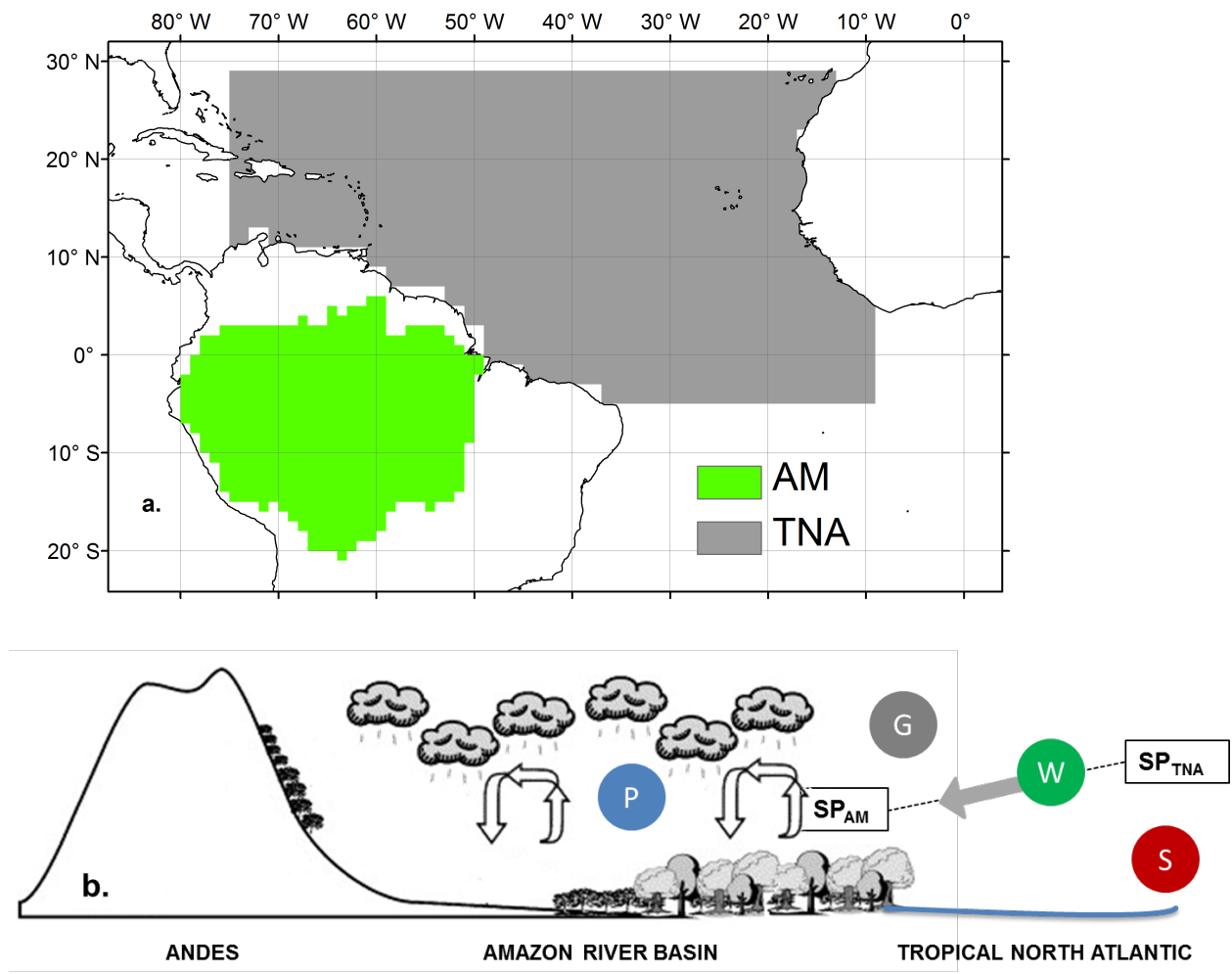

**Fig. 1 a.** Location of the regions under study, with the Amazon River basin (AM) in green, and the Tropical North Atlantic (TNA) in grey. **b**. Illustration of the physical mechanisms and variables involved in the study, *P* is the *P-E* index, *G* is the surface pressure gradient between AM and TNA, *W* represents the zonal winds over the TNA region, and *S* represents TNA SSTs. $SP_{AM}$ and $SP_{TNA}$ are the surface atmospheric pressures on both regions, and the relative position indicates that surface pressure is higher over the TNA than AM. The grey arrow represents the wind flowing between both regions from higher to lower surface atmospheric pressure.

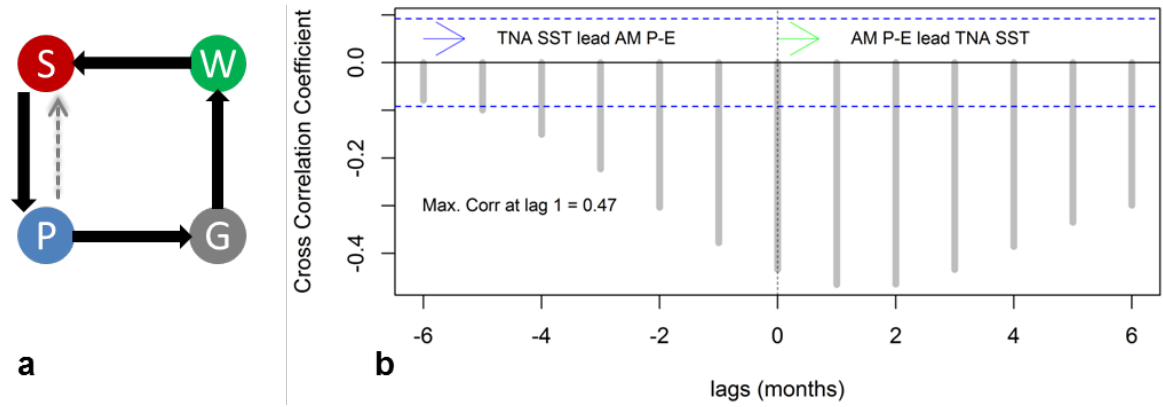

**Fig. 2 a.** Physical mechanisms involved in the AM-to-TNA feedback. Each one of the nodes represents a variable involved in the process: *P* is the *P-E* index, *G* is the surface pressure gradient between AM and TNA, *W* represents the zonal winds over the TNA region, and *S* represents TNA SSTs. The dashed arrow represents the identified lagged cross-correlation found between the *P-E* index and TNA SSTs, the filled arrows represent the sequence of physical processes connecting the Amazon hydrology with TNA SSTs. **b.** Lagged cross-correlations between low-pass filtered SSTs over the TNA and the AM *P-E* index. The Amazon hydrology leads the TNA SSTs, with negative correlations peaking at 1 to 2 month-lags, suggesting that lower (higher) *P* over the AM basin is related with higher (lower) TNA SSTs. The blue dashed lines denote the 95% confidence intervals.

The *P-E* index is defined as the difference between monthly values of Precipitation and Evapotranspiration averaged over the AM River basin, defined from 79.5°W to 50.5°W and from 19.5°S to 4.5°N (Mayorga et al. 2012). We also used time series of zonal wind velocities, *W*, and sea surface temperatures, *S*, over the TNA, and the gradient of surface atmospheric pressures between the TNA and AM regions, *G*.

The precipitation data set was obtained from the Global Precipitation Climatology Centre (GPCC) that produces gridded (1° x 1°) data, based on monthly rainfall collected from more than 80,000 stations worldwide with a rigorous quality control (Schneider et al. 2014). The evapotranspiration data set was obtained from the land surface model ORCHIDEE (1° x 1°), developed and maintained by the *Institute Pierre Simon Laplace* (Krinner et al. 2005). The resulting *P-E* index was calculated from 1979 to 2008, the shared period among both data sets.

The TNA SSTs are defined for the region from 75°W to 10°W and from 5°S to 29°N. The SST data set was obtained from the third version of NOAA's Extended Reconstructed Sea Surface Temperature ERSST (v3b), with a spatial resolution of 1° x 1°, and spanning the period from 1979 to 2008. The SST data has a cold bias correction due to satellite data in previous ERSST reconstructions (Smith et al. 2008). Data sets pertaining to surface atmospheric pressure for both regions (AM and TNA), and zonal wind velocities over the TNA region were obtained from the ERA-Interim Reanalysis (Dee et al. 2011), with a spatial resolution of 0.75° x 0.75°. The analyzed time series were estimated as regional averages, and the surface atmospheric pressure gradient was calculated as the difference in surface pressures between the TNA and AM. We computed the monthly anomalies for each of the series in order to remove the annual cycle. The final time series used in the present study are presented in Fig. 3.

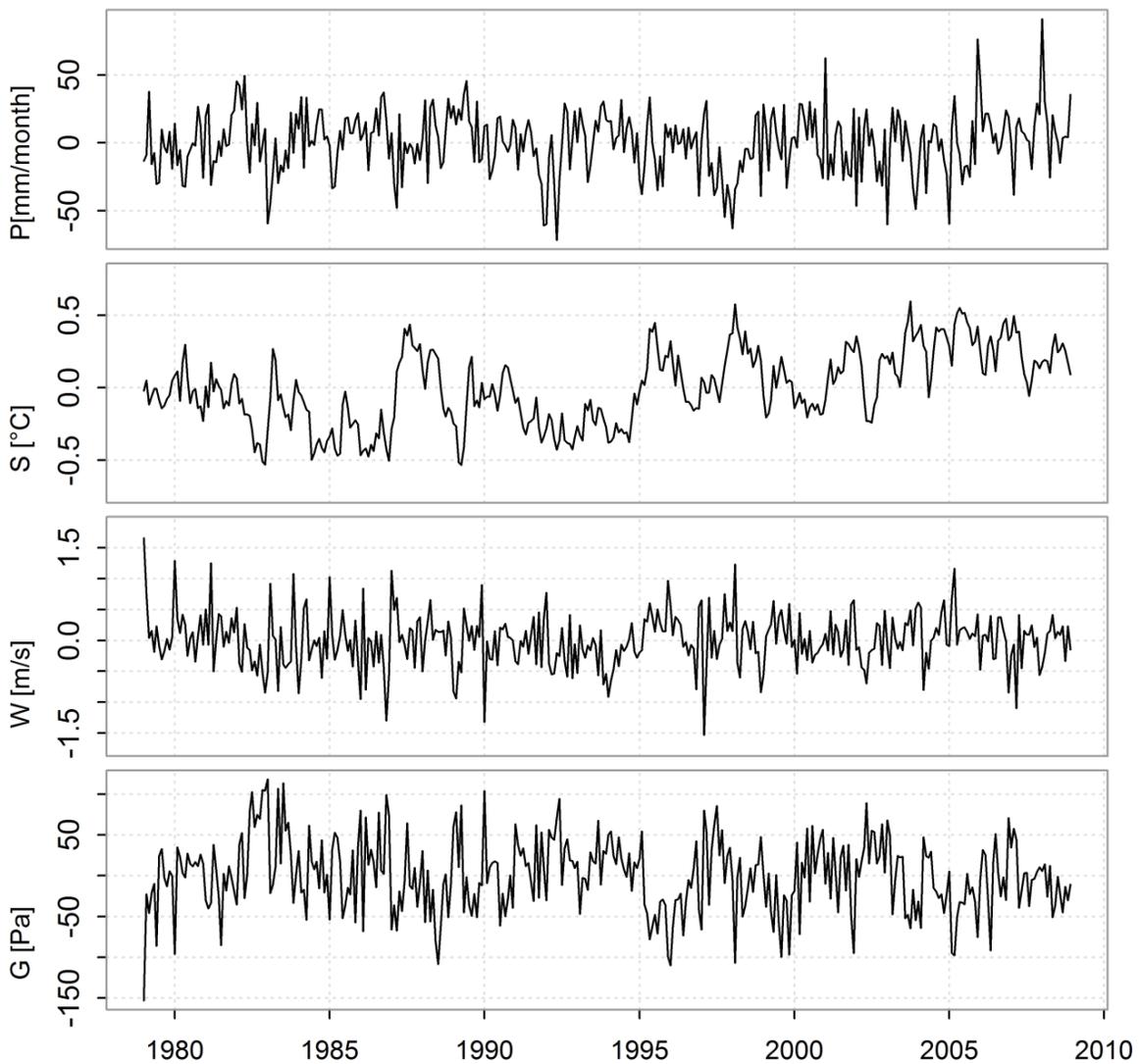

**Fig. 3** Anomaly time series used in the present study, from top to bottom: AM *P-E* index (mm/month), TNA SSTs (°C), TNA zonal wind velocities (m/s), and TNA-AM surface pressure gradient (Pa).

## 3. Methods

### 3.1 Recurrence Analysis

Hydro-climatic and many other dynamical systems tend to have a recurrent behavior in phase space, so that recurrence can be considered as a fundamental property of these kinds of systems (Marwan et al. 2007b). This property can be visualized by the trajectories of the system's dynamics in phase space, and can be quantified by the so-called *Recurrence Plot* (*RP*) (Eckmann et al. 1987). In order to construct an *RP* from a time series, $x_i$, it is necessary to represent the *m*-dimensional phase space of a system *X*. In case of single time series $x_i$, the dynamics has to be artificially reconstructed using the time delay embedding technique (Packard et al. 1980; Takens 1981), whereby the phase trajectories $\vec{x}_i$ are defined as:

$$\vec{x}_i = (x_i + x_{i+\omega}, \ldots, x_{i+\omega(m-1)}), \vec{x}_i \in R^m \quad , (1)$$

where *m* is the embedding dimension and $\omega$ is the time delay. The most common methodologies to quantify both embedding parameters are the false nearest neighbors for the embedding dimension, and the mutual information function for the time-delay (Fraser and Swinney 1986; Marwan 2011). Once the parameters are set, the *RP* can be estimated using the pair-wise proximity test, such that,

$$R_{i,j}^X = \Theta(\varepsilon - \|\vec{x}_i - \vec{x}_j\|) \ i,j = 1, \ldots N', \quad (2)$$

where $N' = N-(m-1)k$ is the number of phase space vectors, $\varepsilon$ is the threshold defined for the proximity between the phase space vectors, $\|\vec{x}_i - \vec{x}_j\|$ is the spatial distance between vectors in phase space, and $\Theta(\ )$ is the Heaviside function: ($\Theta < 0$) = 0, ($\Theta \geq 0$) =1. The plot of the $R^X$ recurrence binary matrix provides the *RP* (Fig. 4). The probability that one system recurs to a certain state $\vec{x}_i$ is equal to the column-average of the recurrence matrix (Marwan et al. 2013):

$$p(\vec{x}_i) = \frac{1}{N'} \sum_{j=1}^{N} R_{i,j}^X \quad (3)$$

Joint recurrence plots (*JRP*) are used to study the possible influence between two physically different systems (Romano et al. 2004; Marwan et al. 2007a), as they provide a measure of the simultaneous recurrence in both. The *JRP* matrix is defined as the Hadamard product of the *RP*s of the single *RP*s of systems *X* and *Y*:

$$JR_{i,j}^{X,Y} = \Theta(\varepsilon - \|\vec{x}_i - \vec{x}_j\|) \times \Theta(\varepsilon - \|\vec{y}_i - \vec{y}_j\|) i,j = 1, \ldots N'. \quad (4)$$

The probability of finding a simultaneous recurrence at time *i* in both systems *X* and *Y* is equal to the column-average of the $JR^{XY}$ matrix:

$$p(\vec{x}_i, \vec{y}_i) = \frac{1}{N'} \sum_{j=1}^{N} JR_{i,j}^{X,Y}. \quad (5)$$

As we are interested in quantifying the dependency between the time series with a specific time lag $\tau$, we will use the log mean recurrence measure of dependence, $RMD(\tau)$, proposed by Goswami et al. (2013), as a measure of the probability that the trajectory of $X$ recurs to the $\varepsilon$ neighborhood of $\vec{x}_i$, when the trajectory of $Y$ recurs simultaneously to the $\varepsilon$ neighborhood of $\vec{y}_i$, after some lag $\tau$, and is given by:

$$RMD(\tau) = log_2 \left( \frac{1}{N''} \sum_{i=1}^{N''} \frac{p(\vec{x}_i, \vec{y}_i(\tau))}{p(\vec{x}_i) p(\vec{y}_i(\tau))} \right) \quad (6)$$

were $N''=N'-\tau$. If both systems $X$ and the time lagged $Y(\tau)$ are independent, then $p(\vec{x}_i, \vec{y}_i(\tau)) = p(\vec{x}_i)p(\vec{y}_i(\tau))$, which implies that $RMD(\tau)=0$. For $\tau > 0$, a non-zero $RMD(\tau)$ implies that $Y$ is dependent on $X$, and the contrary is true for $\tau < 0$.

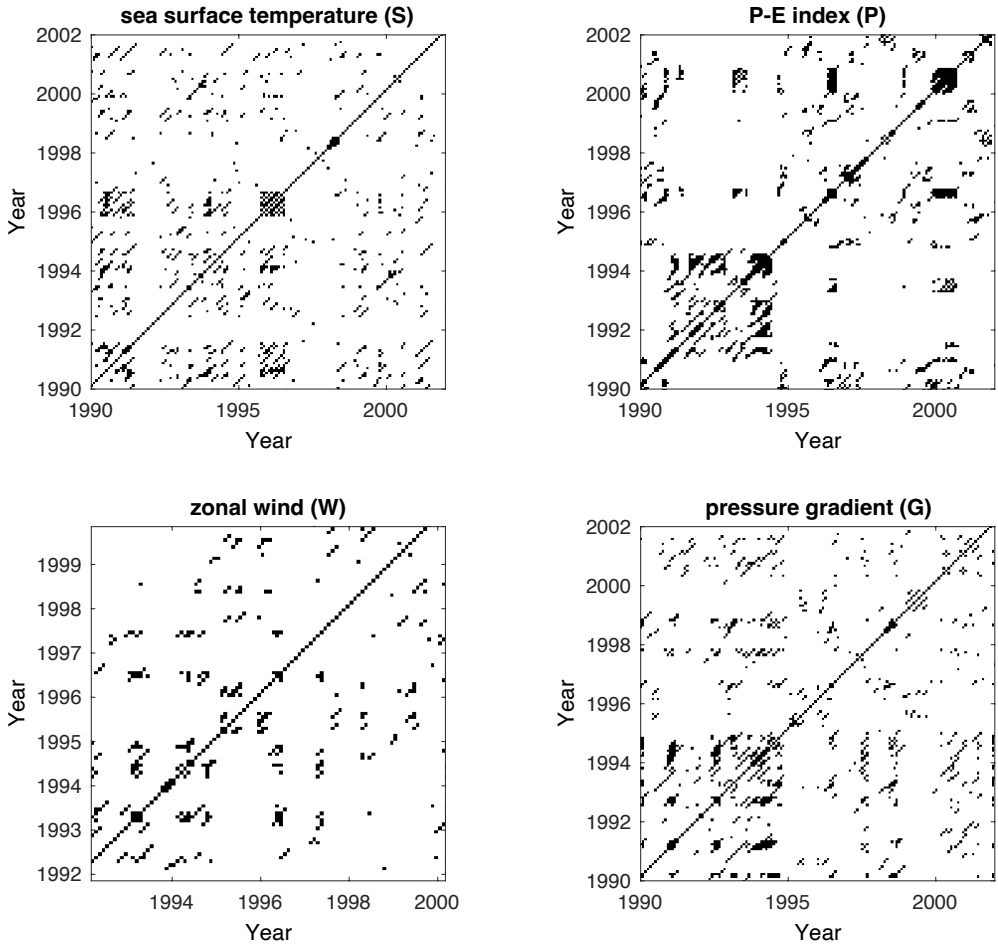

**Fig. 4** Recurrence plots of the time series as presented in Fig. 3, using embedding dimension, $m=4$, time delay, $tau=1$, and a recurrence threshold ensuring a fixed recurrence rate of 5%.

### 3.2 Significance tests

Our approach requires implementing a statistical test for the hypothesis of a joint recurrence between the systems *X* and *Y* at defined time lags. It is based on the joint recurrence between the trajectory in phase space of the original system *X* and a representative number of phase space trajectories of the system *Y*, which represent independent copies of the underlying system, known as twin surrogates (*TS*). For the construction of the *TS* based on the *RP* of a system, we followed the procedure proposed in Thiel et al. (2006, 2008).

Twins can be defined as points in the phase space trajectories that share the same neighborhood up to the threshold $\varepsilon$ such that $R^X_{i,k} = R^X_{j,k}, k = 1, \ldots N$. Thus, the construction of a twin surrogate $\vec{x}^s(t)$ of $\vec{x}(t)$ starts with the identification of all the twin points in a trajectory $\vec{x}$, then one must choose one arbitrary starting point $\vec{x}^s(1) = \vec{x}(k)$. If this $\vec{x}(k)$ has no twin then the next point in the surrogate trajectory is $\vec{x}^s(2) = x(k+1)$, but if $\vec{x}(k)$ does have a twin $\vec{x}(t)$ then the next point can be either $\vec{x}(k+1)$ or $\vec{x}(t+1)$ with equal probability, and thus the process is iterated until the constructed surrogate has the same length of the original time series.

The null hypothesis is that each *TS* trajectory is an independent realization of the system, corresponding to a different initial condition. To test the statistical significance of the *RMD(τ)* measure among *X* and *Y(τ)*, we generate 500 *TS* of the *Y* system, obtain a test distribution of *RMD(τ)* calculated with the observations of the system *X* and each one of the *TS* of *Y*. Finally, we construct the upper 95% and 90% confidence intervals (CI) based on the respective 95th and 90th percentiles of the test distribution. The lags at which values fall outside the confidence band represent the ones where the null hypothesis is rejected, or that the system *X* drives system *Y*, i.e. they are dependent at that particular lag.

## 4. Results and discussion

We used the false nearest neighbor and mutual information methods to estimate the embedding parameters (Fig. A1.1) to construct the joint recurrence plots for the time series studied here. We found $\omega = 1$ and $m = 4$ for *P* and *S*, $\omega = 1$ and $m = 6$ for *W*, and $\omega = 1$ and $m = 5$ for *G*. We decided to set the parameters for all combinations as $\omega = 1$ and $m = 4$ to avoid sparse recurrence plots, as recommended in Marwan (2011). The recurrence threshold, $\varepsilon$, was based on a fixed 5% of the recurrence rate, $RR = 1/N^2 \sum_{i,j} R_{ij}$, for all the time series.

Before presenting the analysis of the dynamical mechanisms involved in the studied feedbacks we will evaluate the recurrence analysis for *P* leading *S*, whose results are shown in Fig. 5. Using the *RMD* measure, we find that the *P-E* index (*P*) in the AM region exerts a significant influence over the TNA STTs from 1 to 2 month-lags, and the *RMD* values are in the 90% confidence band. Comparing this result with the lagged cross-correlations among the two variables presented in Fig. 1b, and the results of (Yoon and Zeng 2010, Fig. 10), one can see that although the relation between the variables is well represented by the two measures, the lagged cross-correlations show a (low) peak at 0 month-lag (simultaneous), while the non-linear measure reflects a higher persistence in time that may last for more than two months.

Fig. 6 shows the recurrence lagged dependence measures of the four time series studied, according to the leading patterns involved in the AM-to-TNA feedback physical processes. Fig. 6a allows us to conclude that *P* is leading *G* by one month lag above the 95% CI, and by zero and two month-lags above the 90% CI. The identified influence of the AM hydrology (mainly precipitation-convection) over the pressure gradient between the two regions at the 0 month lag may be related with the Amazon modulation of the intensity and location of convection in the TNA reported by Wang and Fu (2007) in a time scale from 5 to 7 days.

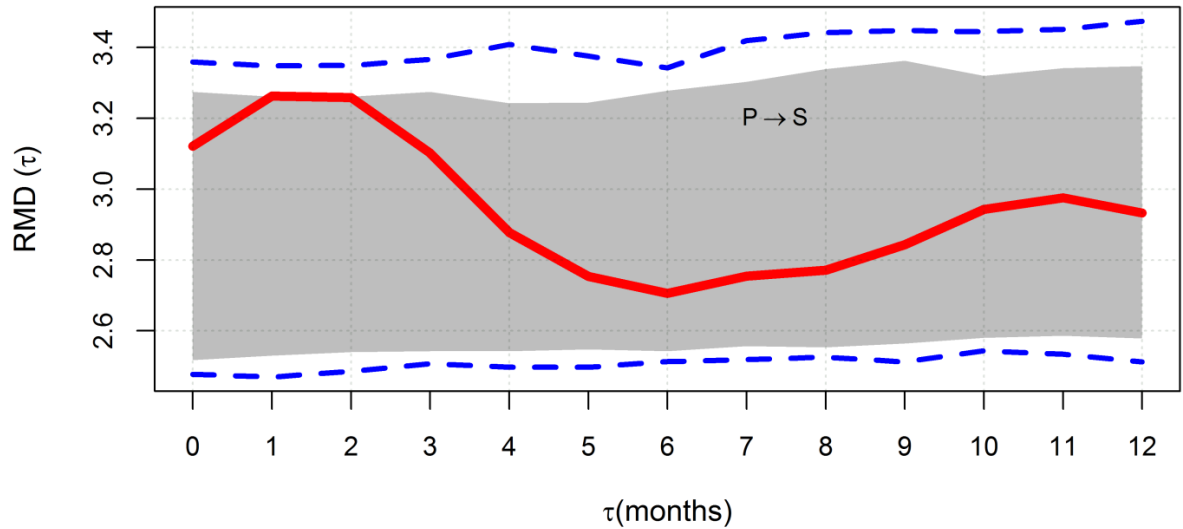

**Fig. 5** Recurrence lagged dependence between the *P* and *S* time series. The arrow denotes the direction of influence between variables; the gray area represents the 90% confidence area, the blue dashed lines represent the 95% confidence intervals, and the red line represents the calculated *RMD* between the variables *P* and *S* (lagged).

Furthermore, *G* (pressure gradient) has a strong influence on *W* (zonal winds) leading from 0 to 3 month-lags, and from 8 to 12 month-lag above the 95% CI, as shown in Fig. 6b. The influence of *G* on *W* may be related with the 30 to 70-days oscillation of tropical winds reported by (Foltz and McPhaden 2008) and the intraseasonal Madden-Julian Oscillation that has important influence in AM (Kayano and Kousky 1999; Garreaud et al. 2009) and over the Atlantic basin, which has been related to the formation of Atlantic tropical cyclones (Klotzbach and Oliver 2015). Finally, *W* leads *S* at 0 to 2 month-lags above the 95% CI as presented in Fig. 6c. The overall interannual results obtained using our non-linear approach confirm the relation between AM and the tropical Atlantic; although for a much larger region over the TNA than the one found by Misra and DiNapoli (2012) regarding a teleconnection between the Amazon and the Tropical Atlantic region at seasonal timescale, mainly forced by wet anomalies in the Amazon, that in turn are associated with cooler SSTs over the Intra-Americas Seas.

According to the results of (Yoon and Zeng 2010), after removal of the ENSO influence on both variables, the relation between AM hydrology and TNA SSTs occurs in phase (lag 0). Using the recurrence analysis we found that the feedback in the AM-to-TNA direction occurs from 0 to 2 month-lags. The whole mechanism that involves the influence of AM land surface-atmospheric processes on the TNA SSTs spans around two months with a 90 to 95% of confidence. This result provides further evidence about the influence of AM hydrological processes over TNA SSTs, and unveils the time span required for convection anomalies over land to affect the oceanic region. These results are also consistent with the empirical analysis of (Poveda and Mesa 1997) regarding the existence of a strong significative statistical association between the hydrology of Andean streamflows leading the TNA SSTs from 3 to 5 months in advance. This result suggests that the hydroclimatological processes acting at continental scale over northern South America and the AM may influence the TNA from one to almost five months.

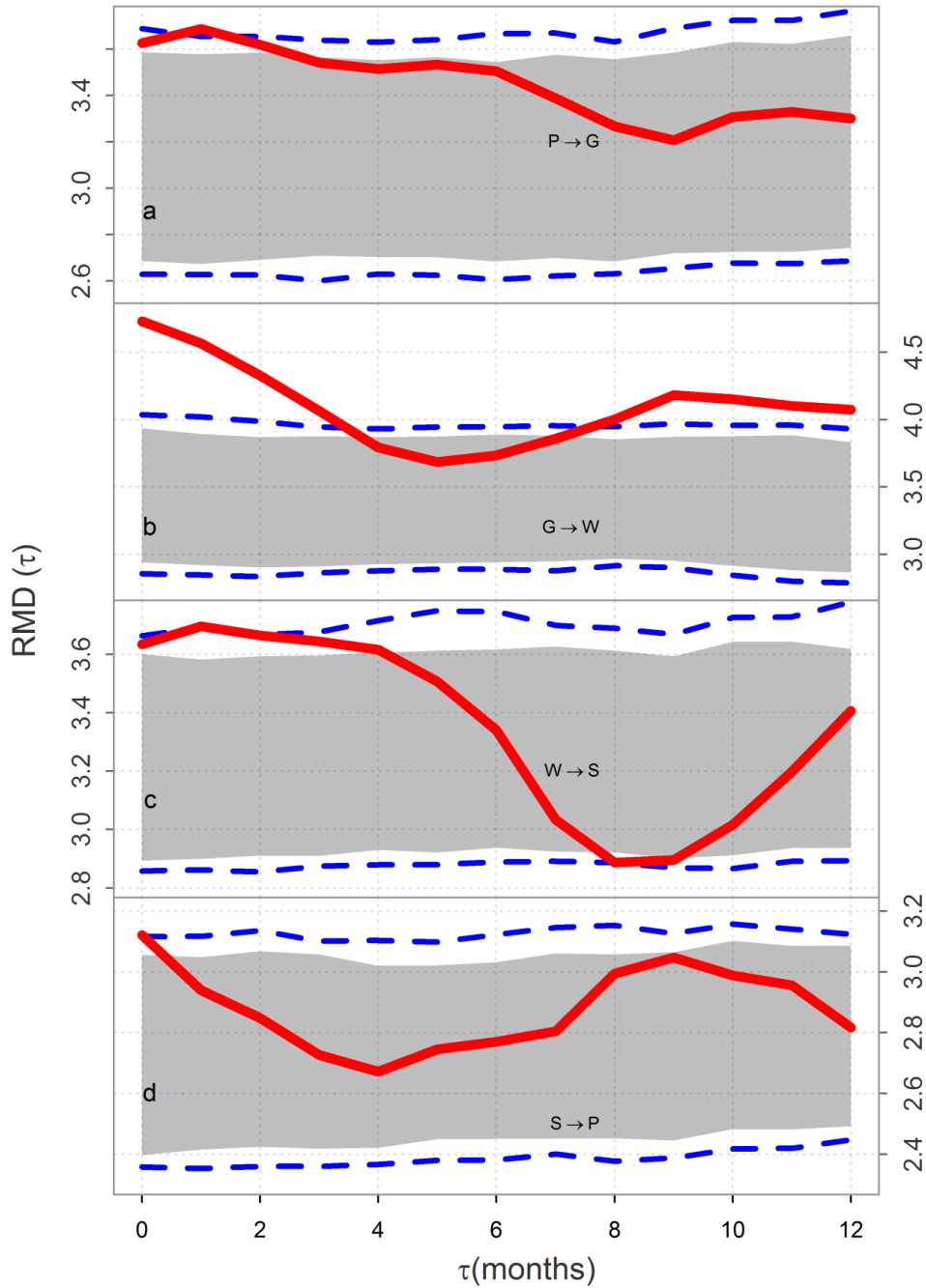

**Fig. 6** Recurrence lagged dependence between the four variables according to the direction identified in processes involved in the AM-to-TNA feedback mechanism. The arrow denotes the direction of influence between the variables, the gray area represents the 90% confidence area, the blue dashed lines denote the 95% confidence intervals, and the red line represents the calculated *RMD* between the variables.

These results further support the existence of two-way feedback mechanisms operating between AM and TNA at interannual timescales. Fig. 6d shows the recurrence analysis of *S* (TNA SSTs) leading *P* (AM hydrology), which point out a simultaneous effect (0 month-lag) with a 95% confidence. We obtained similar results for recurrence rates near 1% (Fig. A1.2). Fig. 7 summarizes the feedback mechanisms and time lags involved in the AM-to-TNA feedback, with the nodes representing the four variables involved.

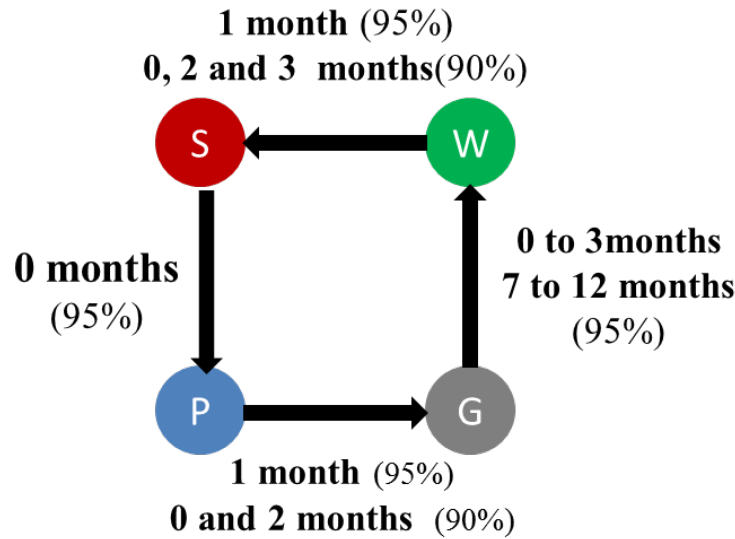

**Fig. 7** Graph representing the studied feedbacks. The numbers in bold represent the time lag of dependency between the variables, and the percentage inside the parenthesis is the level of confidence associated with the identified lags.

Furthermore, our recurrence analyses confirm that two-way feedbacks are set between the two regions at interannual timescales, including a strong influence of AM convection on TNA SSTs, mediated by the surface pressure gradient between the two regions, *G*, and the zonal trade winds, *W*. A summary of the proposed two-way feedbacks mechanism is illustrated in Fig. 8. Furthermore, our results also unveil dependencies between *G* and *W* from 7 to 12-month lags, whose interpretation constitutes a topic of further investigation.

Our work is focused on studying the proposed two-way mechanism from the AM-to-TNA direction. But in order to understand the whole process including the TNA-to-AM direction, we performed a complete recurrence analysis between the variables involved in the studied mechanism. Our results confirm the two-way character of the AM-TNA coupling, as summarized in Appendix 2.

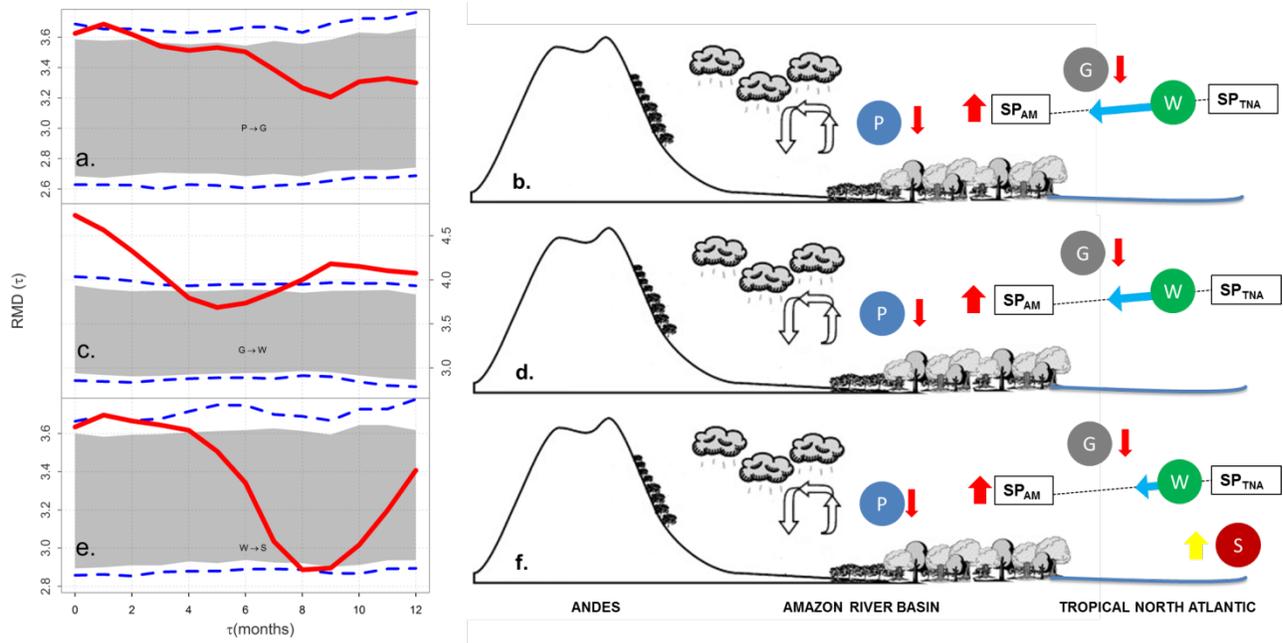

**Fig. 8** Graphical summary of the dynamics of the two-way feedback mechanism originated in the AM region, represented step by step from top to bottom. Left panels show the recurrence analysis results, whereas right panels illustrate the dynamics of the physical processes involved. The arrows beside all variables represent the dynamical changes undergone by the relevant variables. In summary, the mechanism is triggered by a decrease in *P-E* over the AM and the ensuing dynamics develops, as shown in pairs of panels a-b, c-d and e-f. Panel (a) shows the recurrence analysis regarding *P-E* driving *G* from 0-2 months, and panel (b) illustrates that such an influence is mediated by a reduction of *P-E*, an increase of surface pressure in the AM that leads a decrease in *G*. Panel (c) shows the influence of *G* driving *W* from 0-3 months, as discussed in panel (d) through a reduction in the surface pressure gradient between the AM and TNA, and the reduction of the arrow representing the magnitude of wind speed, *W*. The final step in panel (e) shows the influence of *W* over *S* from 0-2 months, through the mechanisms discussed in panel (f), with the appearance of an increase in sea surface temperatures, *S,* due to the ongoing reduction of *W*.

At this point it is necessary to detail the proposed two-way mechanism by disentangling the physical dynamics between the variables *P, G, W* and *S.* We propose that the mechanism is triggered by anomalous convection processes in AM. Therefore, we conduct the physical analysis by evaluating the behavior of monthly anomalies of *P*, *W* and *S* in four particular contrasting extreme convective conditions in the AM, with anomalies greater than ± 20 mm month$^{-1}$. The selected events are those of January 2005 with values of -30 mm month$^{-1}$, and January 2009 with values of +20 mm month$^{-1}$ (Fig.9), September 1999 with values +25 mm month$^{-1}$ and September 2010 with values -25 mm month$^{-1}$ (Fig. 10). The selection of such extreme dry and wet events in the months of January and September will help us to evaluate whether an anomalous interannual increase or reduction in precipitation may trigger the proposed mechanism.

In January of 2005 the AM is experiencing negative anomalies of $P$ of around 30 mm month$^{-1}$ that covers most of the basin (Fig. 9a), with maximum negative anomalies of more than 120 mm month$^{-1}$ in regions over the Andes and central AM. While AM experiences such reductions of $P$ it is possible to observe a reduction in zonal wind velocities from 5°N to 25°N with negative anomalies reaching 1.5 m s$^{-1}$ (Fig. 9b). At the same time, the SSTs between 5°N and 25°N exhibit positive anomalies of around 1.5°C. In January of 2009 the AM is experiencing positive anomalies of $P$ that cover the north-western AM (mean positive anomaly of +20 mm month$^{-1}$), with positive anomalies reaching values of more than 110 mm month$^{-1}$ (Fig. 9d). In the same month it is possible to observe increased zonal wind velocities from 5°N to 30°N, with positive anomalies ranging from 0.5 m s$^{-1}$ to 3 m s$^{-1}$ (Fig. 9e). In phase with the increased zonal wind velocities, SSTs between 5°N and 25°N exhibit negative anomalies of around 0.4°C between 45°E and 20°E (Fig. 9f), and the region of negative SST anomalies moves to 0°N in the subsequent February (not shown). In summary, the response of $S$ due to the influence of $P$ anomalies in the two events analyzed in January occurs from 0 to 2 months-lag. These dynamical evidences during the extreme January events are consistent with the results from the recurrence analysis, which indicate that the complete mechanism from $P$ to $S$ may be triggered and active during 0 to 3 months-lag (Fig. 7).

During September 1999, there are positive anomalies of $P$ over central AM with values around 100 mm month$^{-1}$ (Fig 10a). While AM experiences an increase in $P$, there are two zones of increased zonal wind velocities in the TNA, from 0°N to 10°N, and from 20°N to 30°N there are values of 0.5 m s$^{-1}$(Fig. 10b) and between 10° and 20° there is a zone of reduced zonal winds reaching up to 1.5 m s$^{-1}$. In September 1999, SSTs show an increase of around 0.4°C (Fig. 10c) while during the following month (October 1999), the values reached up to 0.6°C (shown in Fig. 12b) that goes on until November 1999. In September 2010, the entire AM is experiencing negative precipitation anomalies of around 80 mm month$^{-1}$ (Fig. 10d). The zonal wind velocities over the TNA exhibit positive anomalies from 0.5 m s$^{-1}$ to 4 m s$^{-1}$ and from 10°N to 20°N (Fig. 10e). For the same month, values of $S$ exhibit positive anomalies greater than 1°C over the TNA from 5°N to 15°N (Fig. 10f). In summary, for the extreme convection events in September only the one of 2010 lead to a SST warming over the TNA, consistently with the recurrence analysis (Fig. 7). These previous results provide further evidence that convection anomalies over the AM region influence TNA SSTs at interannual timescales.

The two selected extreme events in January and the dry event in September (2010) are consistent with results from our recurrence analysis. For the January events it is possible to observe the configuration of the mechanism following wet and dry convection anomalies in the AM during the Amazon wet season. The unclear development of the mechanism in September may be attributed to the influence of the Atlantic Meridional Mode (AMM) that exerts a stronger influence on the TNA than the one induced by the studied mechanism, and thus anomalous convection in AM during the dry season (such as the ones discussed for September) has less impact on the AM-to-TNA feedback. The AMM, which develops in the boreal spring (MAM) at interannual time scales, is related with the SSTs anomalies over the TNA due to anomalous displacements of the ITCZ, and may be acting in phase with ENSO events magnifying the SST's warming during boreal spring to boreal summer (Mitchell and Wallace 1992; Nobre and Sukla 1996; Good et al. 2008; Foltz and McPhaden 2010; Foltz et al. 2012; Amaya et al. 2016).

Aiming at a more comprehensive identification of the spatial extent of the regions involved in the AM-TNA feedbacks, we define a larger region of the TNA including a broader equatorial portion of the Atlantic and the Intra-American Seas (IAS). From our previous dynamical analysis of the variables *P*, *W* and *S* we may infer that convection over AM might be most related with the TNA region north of 0°N, as is the region experiencing the main changes resulting from the studied mechanism. On the contrary, the more equatorial region of the TNA seems to be more stable and more influenced by the strong seasonality of winds, as proposed by Li and Philander (1997).

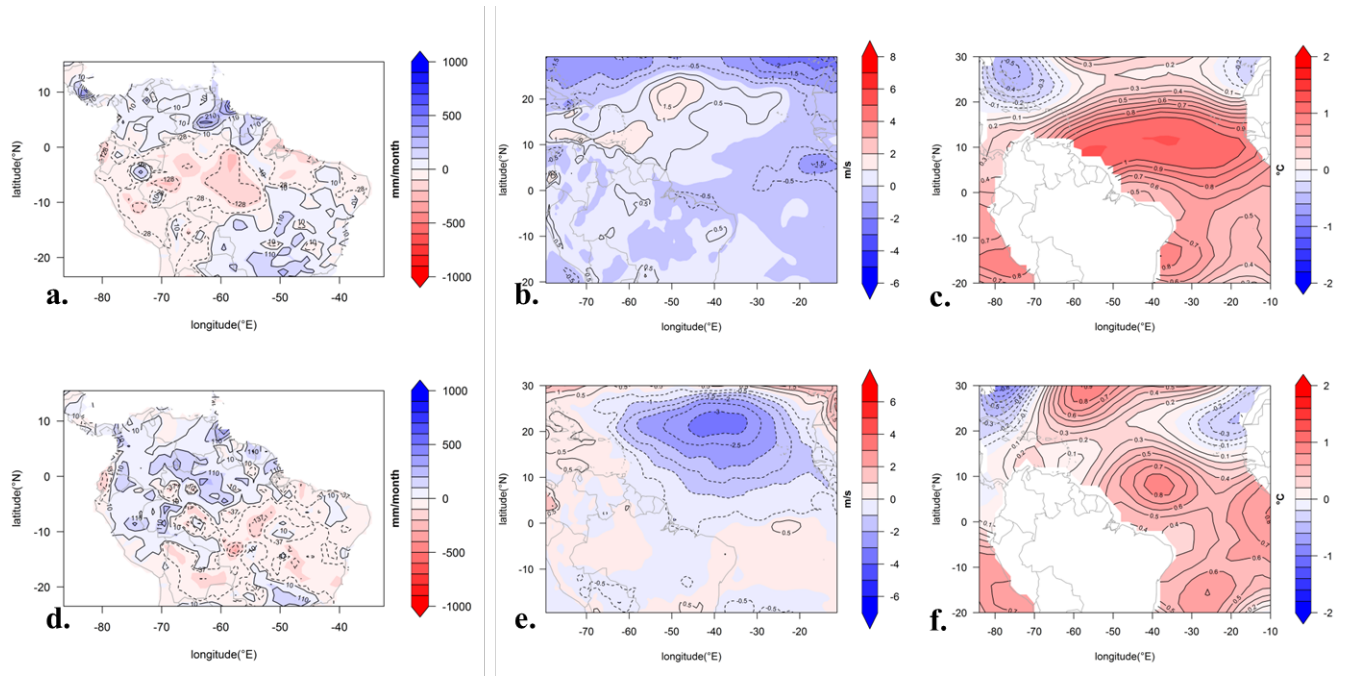

**Fig.9** Monthly anomalies of (a) precipitation, (b) zonal winds, and (c) SSTs for January 2005. Monthly anomalies of (d) precipitation, (e) zonal winds, and (f) SST for January 2009. Dashed (solid) lines denote negative (positive) anomalies.

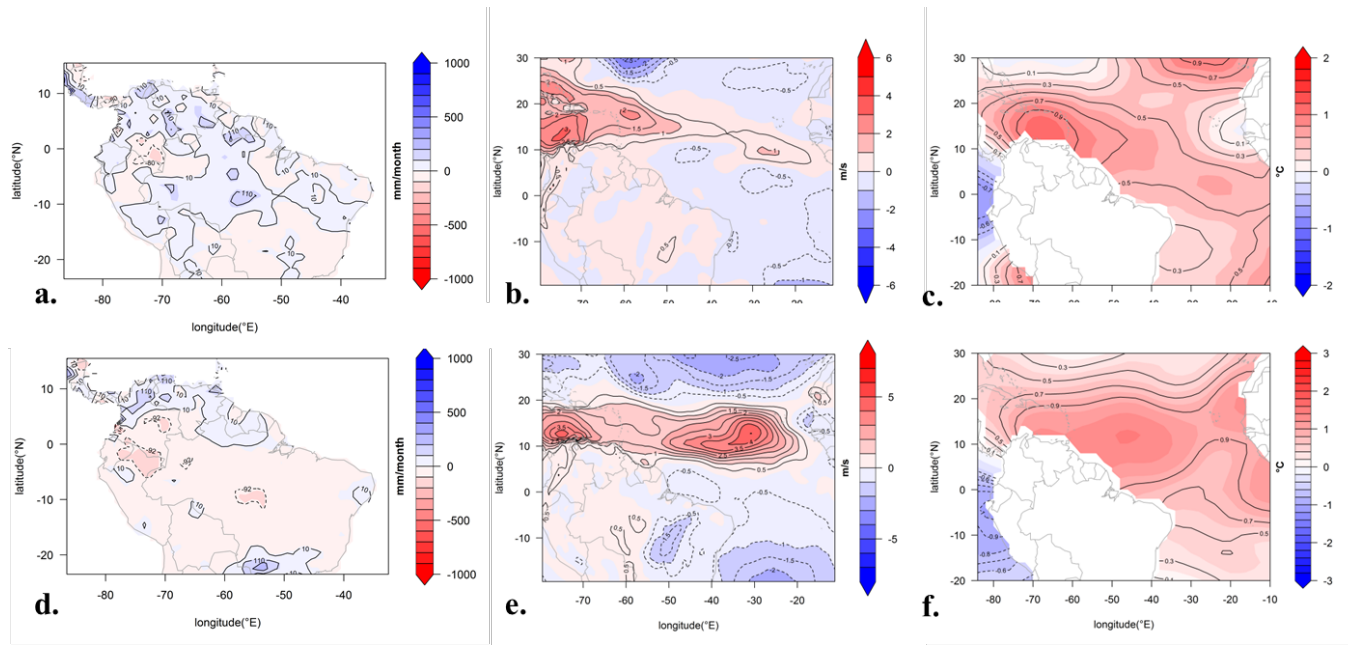

**Fig. 10** Monthly anomalies of (a) precipitation, (b) zonal winds, and (c) SSTs for September of 1999. Monthly anomalies of (d) precipitation, (e) zonal winds, and (f) SST for September of 2010. Dashed (solid) lines denote negative (positive) anomalies.

We also evaluated the monthly mean conditions of the relevant variables during the extreme AM droughts of 1963, 1980, 1983, 1997, 1998, 2005, and 2010, and also during the floods of 1989, 1999, and 2009 (Marengo and Espinoza, 2016). Our aim is to further understand the dynamics of the two-way interactions between AM and TNA in the context of extreme dry and wet events in the AM region, as a key mechanism involved in the warming/cooling of the TNA (Fig. 10). In order to capture the extremes of 1963, 2009 and 2010 that cannot be captured with the *P-E* index (spanning from 1979 to 2008), we used the GPCC average precipitation (*AP*) over the AM to represent hydrological process over land.

During the extreme AM droughts (red lines in all panels of Fig. 11), average precipitation (*AP*) in January reaches negative anomalies of around 20 mm month$^{-1}$, thus implying a reduction of convective processes during the peak of the wet season. One month later, in February, the surface atmospheric pressure gradient between AM and TNA (*G*) reaches a minimum value, in phase with minimum zonal wind velocities (*W*) from the TNA towards the AM. During the first three months of the year, there is a sustained SST warming that continues until May. The further warming of the SSTs from April to May, after the influence of the reduced convection in AM, may be related to the effects of the AMM during boreal spring that induces warmer SSTs at interannual timescales. On the contrary, during extreme AM floods (blue lines in all panels of Fig. 11) there is an increased precipitation in January followed by higher pressure gradients, faster winds towards the AM, and cooler TNA SSTs. As suggested in our previous results convection anomalies may trigger the mechanism, but during the anomalous peak of convection found in June in flood years (see *AP* in Fig. 11) there is no evidence of a concomitant increase in *G*, which may be explained by the control that the ongoing SST warming from April to August exerts over the oceanic surface pressure, and thus over *G*. Although the 0 to 3 month-lag found in the AM-to-TNA feedback mechanism is well represented in the behavior of the time series of mean conditions during extreme events in the AM (light blue shadings in Fig. 11), wet and dry years in the AM have different dynamics and the action of the mechanism is most evident during periods of negative precipitation anomalies in January.

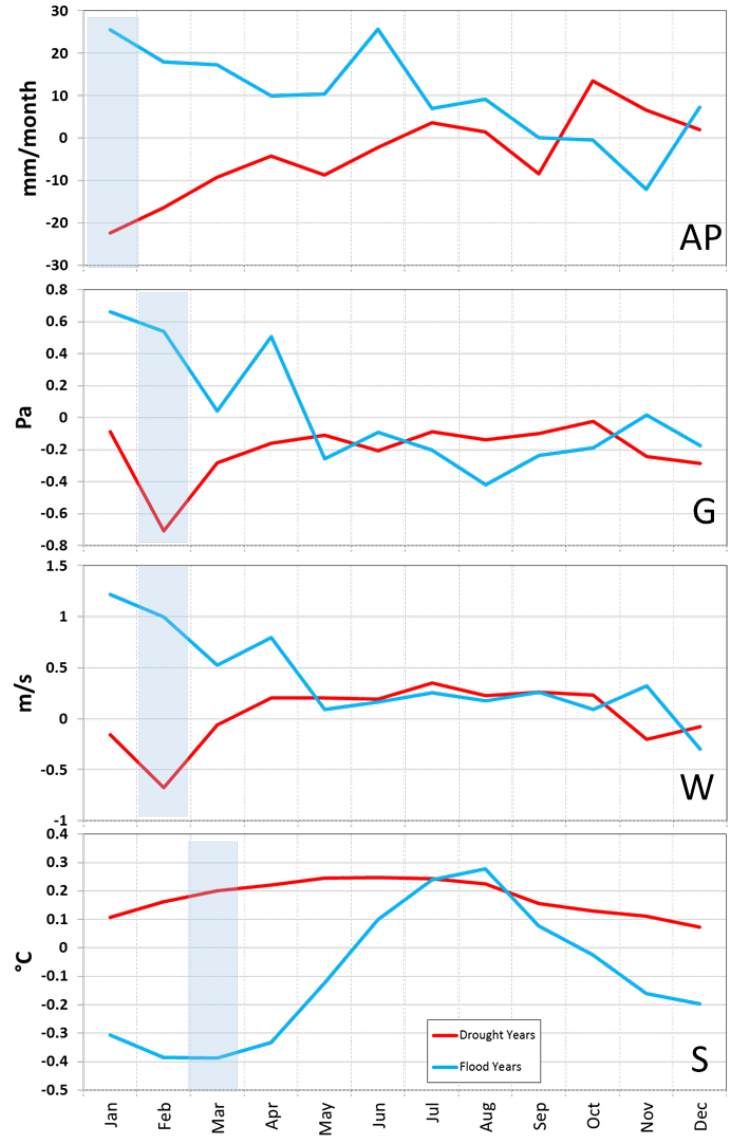

**Fig. 11** Monthly mean values of GPCC precipitation (*AP*), surface atmospheric pressure gradient between AM and TNA (*G*), zonal winds (*W*), and SSTs (*S*), during extreme dry years (red lines): 1963, 1980, 1983, 1997, 1998, 2005, and 2010, and during extreme wet years (blue lines): 1989, 1999, and 2009 in AM. Negative zonal winds represent a reduction of the winds flowing from the TNA to the AM. Light blue shading represents the temporal progression of the mechanism during dry years from the begging of the year (peak of the AM wet season).

So far we have shown the physical mechanisms that connect the AM and TNA regions at interannual timescales, specifically during extreme events of precipitation in the AM. The influence of AM during such events may be mediated by a common driver such as the ENSO system. To verify the influence of ENSO in the two-way feedback mechanisms between AM and TNA, we compare the time evolution of the variables during four selected extreme events, defined in Table 1 of Marengo and Espinoza (2016): the flood of 1999 related to La Niña, the drought of 2010 related with El Niño, and two other events unrelated to ENSO: the 2009 flood, and the 2005 drought. Fig. 12 shows the time evolution of the relevant variables for the chosen extreme years. The evolution of the mechanism during the droughts (Fig. 12 a) indicates that in both episodes *AP* exhibits a strong negative anomaly in January, followed by negative anomalies in *G* and *W*, and warmer than normal SSTs; both droughts showing differences in the magnitude and behavior of the precipitation anomalies, which implies that the mechanism operates differently: for the non ENSO-related drought of 2005 (Fig. 12a, left) there is a delay of two months between the peaks of *AP* and *G*, and of one month between *W* and *S*. In the case of the ENSO-related drought of 2010 (Fig. 12a, right) the decrease in *G* and *W* occurs simultaneously, and with a 2-month delay between *W* and *S*. It is worth noticing that *G* exhibits abrupt changes in March 2010, which may be related to the influence of ENSO in the surface pressure of both regions due to an anomalous Walker circulation in the Amazon (Lewis et al. 2011; Marengo et al. 2011).

For the flood years (1999 and 2009) the peaks of *AP, G* and *W* occur simultaneously (Fig. 12a and Fig. 12b), and the whole mechanism is fully fledged from January to February. This result is explained given that from November onwards the Amazon River basin is experiencing the wet season, and thus positive precipitation anomalies are present even before January. For the 2009 flood (Fig. 12b, left) one may expect that the *AP* peak of March will trigger the mechanism, but the ocean is experiencing the passage of the ITCZ, so there is another forcing in *S* that influences also the evaporative cooling in the ocean. These results may indicate that although the mechanism is well developed during the occurrence of El Niño and La Niña episodes, it is also present when the extreme convection events in the AM are not influenced by ENSO. There are two peaks of anomalous precipitation during the months of September 2010 (Fig. 12 a) and September 1999 (Fig. 12b) that may be related with the triggering of the mechanism. During 2010 the time evolution of the variables is not conclusive, and during 1999 the mechanism appears to be weak and to take three months from the *P* anomaly to the SST anomaly. An extensive analysis was already presented in the discussion of Fig. 10.

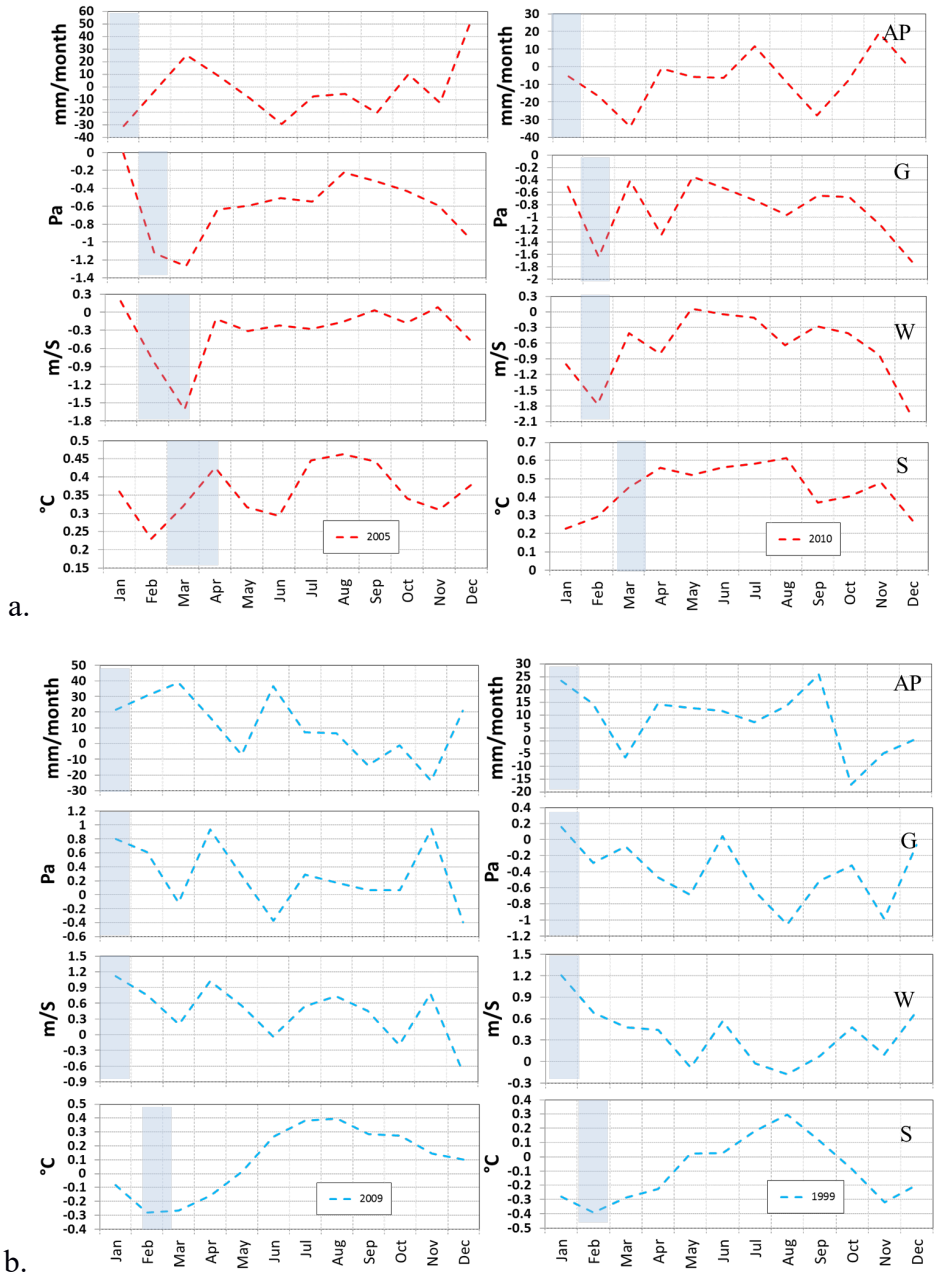

**Fig. 12** Average GPCC precipitation (*AP*), surface atmospheric pressure gradient (*G*), zonal winds (*W*), and SSTs (*S*), for the droughts of 2005 and 2010 (a), and the floods of 1999 and 2009 (b) in AM. ENSO-related extremes depicted in the panels of the right in (a) and (b), and non-ENSO related extremes depicted in the left panels in (a) and (b). Negative zonal winds represent a reduction of the winds from the TNA towards the AM.

## 5. Concluding remarks

Convective processes in the Amazon are characterized by a strong seasonality, where climate systems such as ENSO (interannual) and the migration of the ITCZ (seasonal) are key players. The moisture flux through the Amazon River basin and further south in the continent is driven by the well-known development of the SAMS system and the important source of moisture from the TNA to the AM. Thus the availability and dynamics of moisture in the Amazon are of utmost importance for the stability of the whole hydroclimate in South America. There is an ongoing warming of the TNA SSTs, reported as a part of a decadal cycle of the Atlantic that has a strong relation with droughts in the Amazon (Lewis et al. 2011; Marengo et al. 2011). Also, in recent years there have been reports on how the surface and atmospheric branches of the hydrological cycle in the Amazon River basin exhibit a positive trend toward wetter conditions (Gloor et al. 2013) and how the dry season tends to lengthen in duration (Fu et al. 2013).

Therefore, the identified two-way feedback mechanisms between the AM and TNA regions are of paramount importance towards understanding the dynamics of particular of extreme events at interannual timescales over both regions. Our results point out a significant influence of the AM hydrology over the TNA SSTs, and unveil that such influence is complex and has to be mediated by the surface atmospheric pressure gradient between both regions and by the zonal winds ($G$ and $W$). We have also found that the identified feedback processes develops from 0 to 2 months after convection anomalies in the AM, and that dry conditions in the AM have the potential to increase TNA's SSTs.

Using a nonlinear technique, we have gone further in the analysis of the teleconnections between the AM and the TNA, putting forward the hypothesis that convective negative (positive) anomalies in the AM can trigger an increase (decrease) of the TNA SST's not necessarily mediated by ENSO. Our results allow us to conclude the AM is not a passive spectator of large scale ocean-atmopsheric phenomena, but a key player in the tropical climate at interannual time scales. We have also found that the AM influence over the TNA at interannual timescales goes beyond the IAS region (Misra and DiNapoli, 2012), covering an even greater extent of the Tropical Atlantic Ocean. The dynamical analysis strongly supports our results from the recurrence analysis regarding the coupling of the studied variables in time.

There is evidence suggesting that the Amazon interannual hydrological variability is getting more extreme (Fu et al. 2013; Gloor et al. 2013; Yin et al. 2014; Zou et al. 2015). According to our results, a continuous warming of the TNA SST's may be amplified by a drier Amazon, and may affect the region's ocean-atmosphere dynamics and diverse mechanisms and processes such as tropical easterly waves, tropical storms, which might induce more frequent and intense hurricanes over the TNA and the Caribbean Sea, that in turn could cause positive feedbacks into the continent, thus accelerating desiccation and vegetation dieback of Amazonia, which in turn could lead to a tipping point in the Amazonian hydroclimate (Lenton et al. 2008), with significant consequences at continental and global scales.

As suggested by our results, future anomalies in rainfall over the Amazon River basin possibly caused by climate change, climate variability, deforestation and land use/land change might thus be related to anomalous future stages of TNA's SSTs. Our results also suggest that the Amazon plays a key role in the TNA warming, reinforcing the feedback and triggering more severe droughts, such as the well-studied two one-in-a-century record breaking events of 2005 and 2010 in the Amazon.

The *RMD* results presented in this study points out the time frame in which the proposed mechanism develops, which is evident in some of the extreme events analyzed by means of mapping anomaly fields. Also there are extreme events where the mechanism is not detected, and we can conclude that the non-linear tool may be reflecting the average behavior of the mechanism and/or that the proposed mechanism may be acting together with other forcing like AMM. Therefore *RMD* have proved to be a useful tool for climate variability analysis and our results may be considered as a stepping stone for detailed modeling studies that deepens in the dynamics of the mechanism while isolating other sources of variability in the region.

Although the recurrence measure reported here has proven to be useful in the study of lagged dependencies, there are still shortcomings that can be addressed like the selection of parameters or the identification of the type of interaction between the variables in order to detect positive or negative dependence. Other dynamical systems measures as the transfer entropy is an alternative to quantify causality between variables excluding the presence of an extra common driver, although it requires longer time series and greater computational resources.

**Acknowledgments**

Alejandro Builes-Jaramillo was partially supported by the program "Research Grants - Short-Term Grants, 2015 (57130097)" of the Deutscher Akademischer Austauschdienst (DAAD) and by the Humboldt University of Berlin. The contribution of Norbert Marwan was supported by the project DFG RTG 2043/1 Natural hazards and risks in a changing world. The work of G. Poveda was supported by Universidad Nacional de Colombia at Medellín, as a contribution to the AMAZALERT research programme, funded by the European Commission. Recurrence analysis was carried out with the CRP Toolbox for MATLAB developed by Norbert Marwan and available at http://tocsy.pik-potsdam.de/CRPtoolbox.

**Appendices**

**Appendix 1. Recurrence Technique**

From the mutual information analysis we selected the lag value that corresponds to the one where mutual information steep changes to a slower decrease in its steep as the optimum value for $\tau$ from Fig. A1.1 we observe that for the P-E index $\tau =1$, also Fig. A1.1 shows that the value of $m$ that corresponds to the lowest percent of false neighbors in the reconstructed phase-space is 4. The same procedure was used for all the time series (not shown).

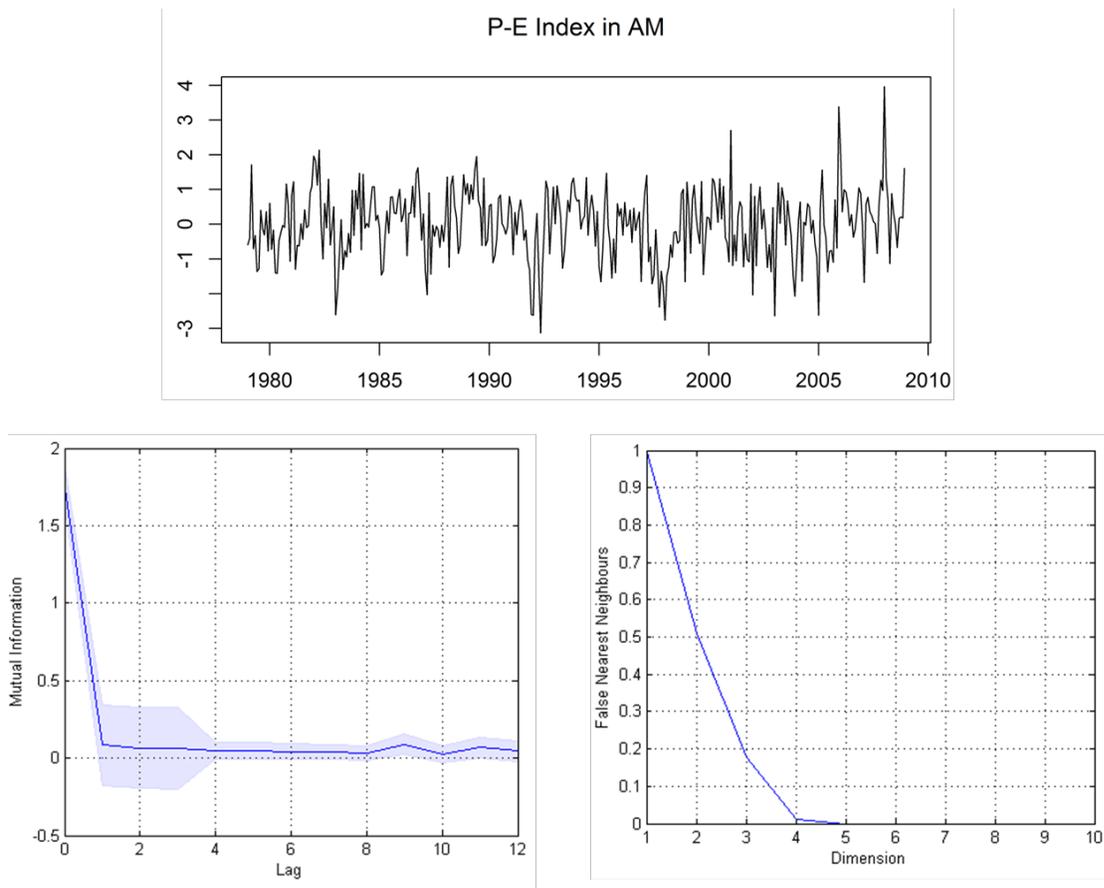

**Fig. A1.1** Mutual information with respect to time delay $\tau$ and the false nearest neighbors with respect to the dimension $m$ for the AM P-E index.

The lagged dependence analysis for a 1% of the recurrence rate maintains the same results previously described for the feedback mechanism between the AM and TNA regions, and in some of the lags increasing the confidence of the results, this increase in confidence for some of the lags may be a result of the finer threshold that cleans spurious proximities between the trajectories and that may reduce the effect of noise in the time series (Fig. A1.2 a, b and c).

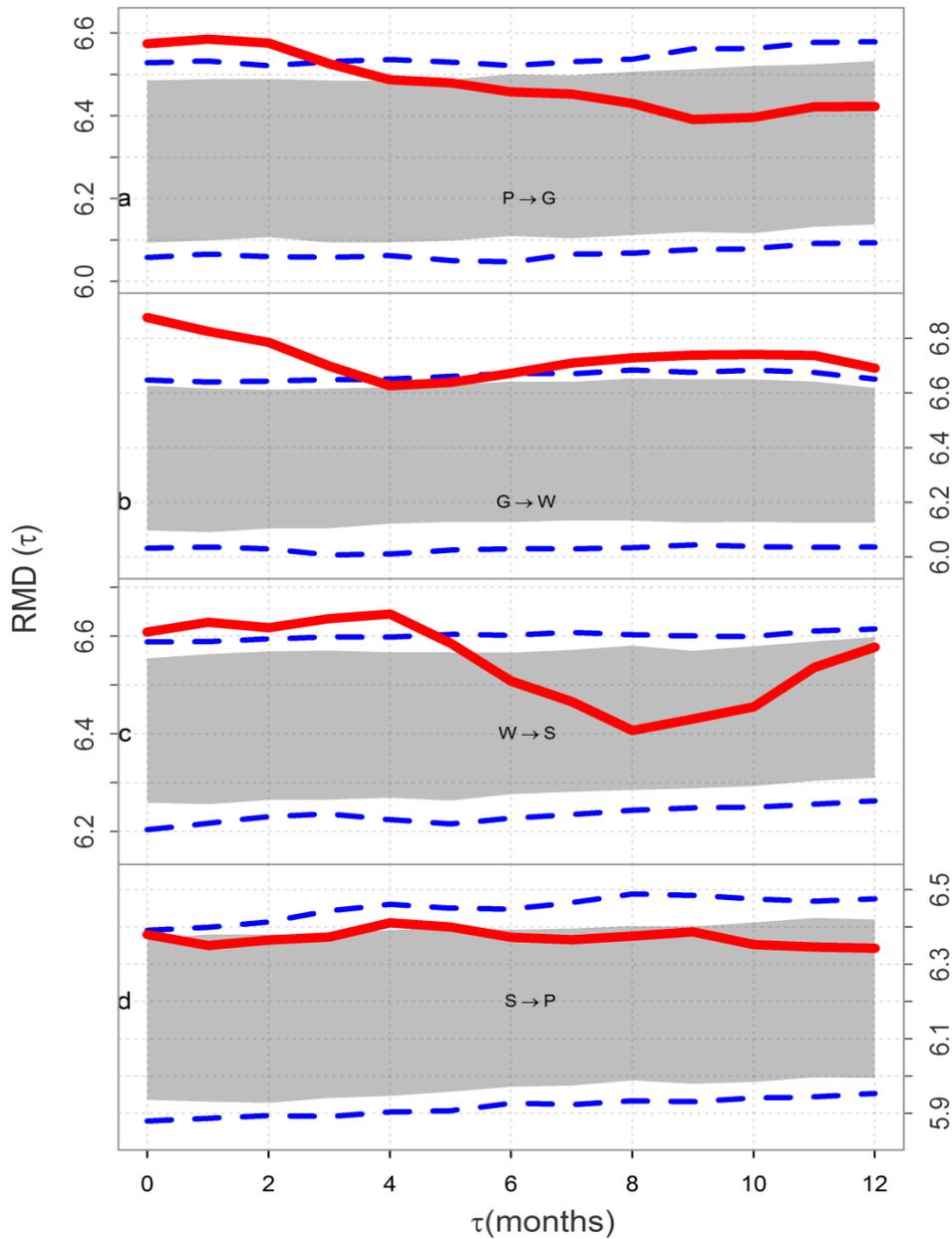

**Fig. A1.2** Recurrence lagged dependence between the four variables according to the direction defined in the mechanism of feedback. The arrow between the names of the variables denotes which is the leading variable, the gray area represents the 90% confidence area, the blue dashed lines represent the 95% confidence intervals and the red line represents the calculated RMD between the variables. The recurrence threshold $\varepsilon$ was based on a fixed 1% of the recurrence rate $RR = 1/N^2 \sum_{i,j} R_{ij}$ for all the time series.

## Appendix 2. Complementary recurrence analysis

We compute a complementary recurrence analysis of the two-way mechanism, and analyze the dependence between the variables in both trajectories. In Fig. A2.1 we illustrate how the two-way feedback mechanism operates among the variables involved in process (such plot is furthered explained in Fig. 6 of the manuscript). The proposed two-way mechanism is supported by the significant correlations between the variables for lags from -2 to 2 months. Such results confirm that the mechanism acts as a two-way process, such that the TNA STTs affects the AM hydrology and vice versa. The recurrence results supports the well-known influence of the TNA SSTs on the surface winds S->W (Chung et al. 2002) is presented in Fig. A2.1.

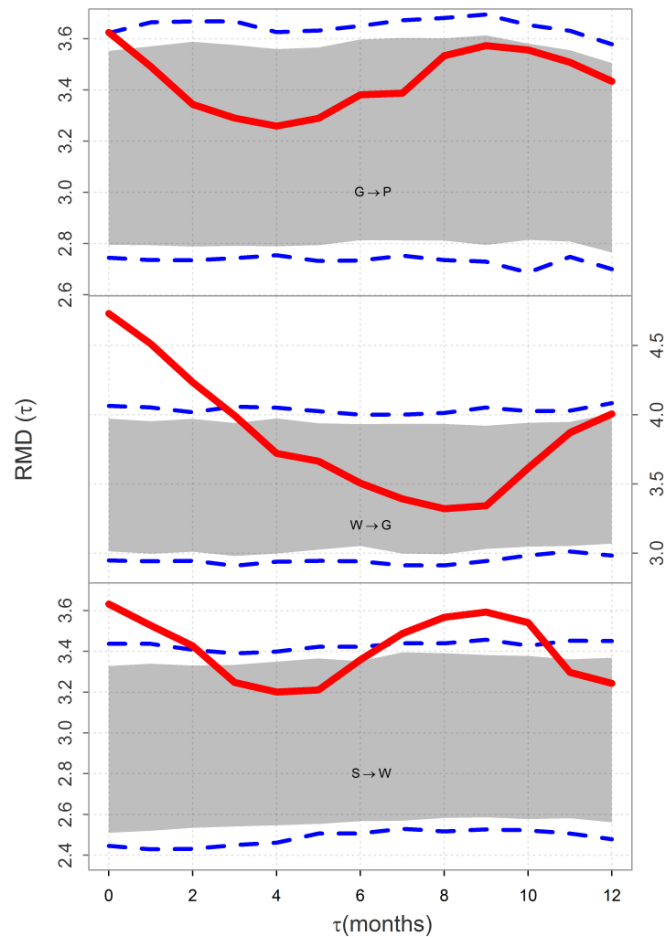

**Fig. A2.1** Recurrence lagged dependence between G -> P, W -> G and S -> W. The arrow denotes the direction of influence between variables, the gray area represents the 90% confidence area, the blue dashed lines denote the 95% confidence intervals, and the red line represents the calculated RMD between variables. The recurrence threshold $\varepsilon$ was based on a fixed 5% of the recurrence rate $RR = 1/N^2 \sum_{i,j} R_{ij}$ for all the time series, as in Fig. 6.

In Fig. A2.2 we present the two-way relations between the variables G and S, as well as P and W. Once the surface winds are affected by the TNA SSTs, then the pressure gradient (G) between the TNA and AM is affected. Cooling or heating of the TNA SSTs are also related with changes in G. The relation S<->G can be seen also as a two-way relationship where the SSTs drive the atmospheric pressure gradient for several months (0 to1, and 4 to 8) while the pressure gradient drives the SSTs during a period of 2 months (0-2). According to the recurrence analysis P does not seem drive W, and therefore any connection between those two variables over the AM and the TNA is to be mediated by other variables in a nonlinear way as evidenced in the recurrence analysis. The influence of W over P is significant during the entire year confirming the well-established fact that there is a direct influence of zonal winds in the transport of moisture from the ocean to the continent to influence convective process in the AM (Yoon and Zeng 2010; Moraes-Arraut et al. 2011; Poveda et al. 2014)

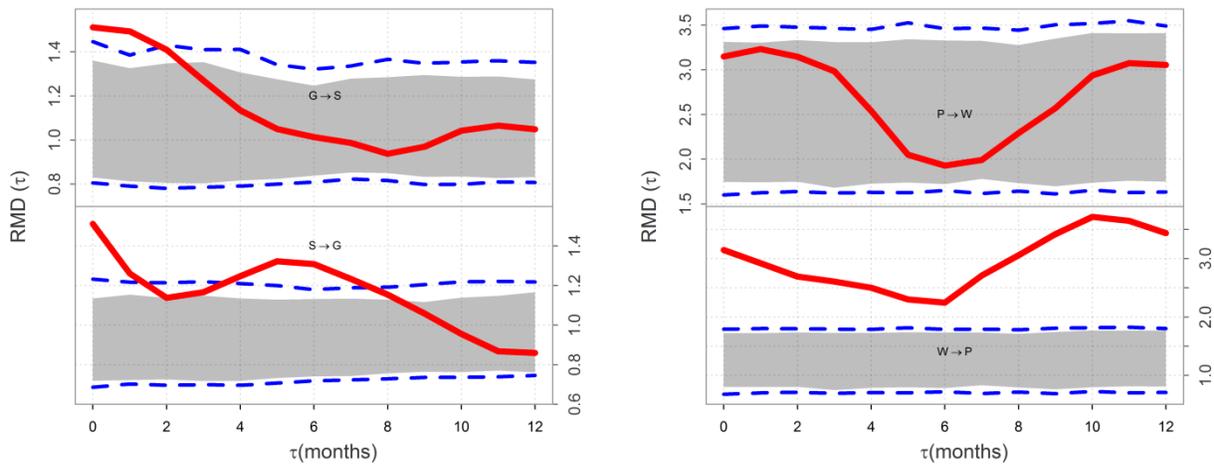

**Fig. A2.2** Recurrence lagged dependences supporting the two way influences between G < - > S (left panel) and P < - > W (right panel). The arrow denotes the direction of influence between the variables, the gray area represents the 90% confidence area, the blue dashed lines denote the 95% confidence intervals, and the red line represents the calculated RMD between variables. The recurrence threshold ε was based on a fixed 5% of the recurrence rate $RR = 1/N^2 \sum_{i,j} R_{ij}$ for all the time series, as in Fig. 6.
32